\documentclass[8.5pt,twoside,twocolumn]{article}
\usepackage[super,sort&compress,comma]{natbib}
\usepackage{mhchem}
\usepackage{times,mathptmx}
\usepackage{sectsty}
\usepackage{balance}

\usepackage{graphicx} 
\usepackage{lastpage}
\usepackage{amsmath}
\usepackage{amssymb}
\usepackage[format=plain,justification=raggedright,singlelinecheck=false,font=small,labelfont=bf,labelsep=space]{caption}

\begin{document}

\thispagestyle{plain}
\renewcommand{\thefootnote}{\fnsymbol{footnote}}
\renewcommand\footnoterule{\vspace*{1pt}%
\hrule width 3.4in height 0.4pt \vspace*{5pt}}
\setcounter{secnumdepth}{5}

\makeatletter
\def\subsubsection{\@startsection{subsubsection}{3}{10pt}{-1.25ex plus -1ex minus -.1ex}{0ex plus 0ex}{\normalsize\bf}}
\def\paragraph{\@startsection{paragraph}{4}{10pt}{-1.25ex plus -1ex minus -.1ex}{0ex plus 0ex}{\normalsize\textit}}
\renewcommand\@biblabel[1]{#1}
\renewcommand\@makefntext[1]%
{\noindent\makebox[0pt][r]{\@thefnmark\,}#1}
\makeatother
\renewcommand{\figurename}{\small{Fig.}~}
\sectionfont{\large}
\subsectionfont{\normalsize}

\setlength{\arrayrulewidth}{1pt}
\setlength{\columnsep}{6.5mm}
\setlength\bibsep{1pt}

\twocolumn[
  \begin{@twocolumnfalse}
\noindent\LARGE{\textbf{Interactions of neutral semipermeable shells in asymmetric electrolyte solutions}}
\vspace{0.6cm}

\noindent\large{\textbf{Vladimir Lobaskin,$^{\ast}$ \textit{$^{a}$} Artem N. Bogdanov,\textit{$^{b}$} and Olga I. Vinogradova\textit{$^{b,c,d}$}}
}\vspace{0.5cm}

\noindent \textbf{\small{DOI: 10.1039/C2SM25605C}}
\vspace{0.6cm}

\noindent \normalsize{We study the ionic equilibria and interactions of neutral semi-permeable spherical shells immersed in electrolyte solutions, including polyions. Although the shells are uncharged, only one type of ions of the electrolyte can permeate them, thus leading to a steric charge separation in the system. This gives rise to a charge accumulation inside the shell and a build up of concentration-dependent shell potential, which converts into a disjoining pressure between the neighboring shells. These are quantified by using the Poisson-Boltzmann and integral equations theory. In particular, we show that in case of low valency electrolytes, interactions between shells are repulsive and can be sufficiently strong to stabilize the shell dispersion. In contrast, the charge correlation effects in solutions of polyvalent ions result in attractions between the shells, with can lead to their aggregation. }

\vspace{0.5cm}
 \end{@twocolumnfalse}
  ]

\section{Introduction}

\footnotetext{\textit{$^{a}$~School of Physics and Complex and Adaptive Systems Lab, University College Dublin, Belfield, Dublin 4, Ireland. E-mail: vladimir.lobaskin@ucd.ie}}
\footnotetext{\textit{$^{b}$~A.N. Frumkin Institute of Physical Chemistry and Electrochemistry, Russian Academy of Sciences, 31 Leninsky Prospect, 119071 Moscow, Russia.}}
\footnotetext{\textit{$^{c}$~Department of Physics, M.V. Lomonosov Moscow State University, 119991 Moscow, Russia. }}
\footnotetext{\textit{$^{d}$~DWI, RWTH Aachen, Forckenbeckstr. 50, 52056 Aachen, Germany. }}

In recent years, there has been much interest in creating
self-assembled micro- and nanocontainers for various technological applications, including pharmaceutics, cosmetics, food, chemical and biotechnologies. In contrast to typical
macroscopic containers, the shells of microcapsules are molecularly thin and, therefore, it is usually hard if not
impossible to make them completely impermeable for the solvent, the small ions, or low molecular weight solutes. Typical examples
of such shells are vesicles and liposomes with ionic channels,\cite{menger.fm:1998,lindemann.m:2006} various types of
micro- and nanocapsules,\cite{donath.e:1998,vinogradova.oi:2006,vinogradova.oi:2004b,kim.bs:2007}
cell\cite{alberts.b:1983,darnell.j:1986} and bacterial\cite{sen.k:1988,stock.jb:1977,sukharev.s:2001}
membranes, viral capsids.\cite{cordova.a:2003,odijk.t:2003} Such shells in a contact with a polyelectrolyte solution have been employed for measuring elastic\cite{gao.c:2001,vinogradova.oi:2004,vinogradova.oi:2004b} and elasto-plastic\cite{kim.bs:2011} modulus of the
shell's material and for selective encapsulation.\cite{lvov.y:2001,lebedeva.ov:2005,kim.bs:2005d}
Recent studies have shown that mechanical properties of semi-permeable capsules filled with polyelectrolytes, their aggregation and
and sedimentation stability are strongly influenced by ionic Donnan equilibria
between the container interior and the surrounding solution.\cite{vinogradova.oi:2006,kim.bs:2007}

During the last few years several theoretical and simulation papers have been concerned with the Donnan equilibria in charged systems with semi-permeable walls.
A large fraction of these deal with the calculation of ion profiles and the excess osmotic pressure on the membrane, by using the \emph{linearized} mean-field Poisson-Boltzmann (PB) theory\cite{deserno.m:2002,tsekov.r:2006,stukan.mr:2008,note2} and molecular dynamics simulations.\cite{stukan.mr:2006,stukan.mr:2008} Other authors used integral equation theories to address the effect of the wall thickness on the electrolyte distribution.\cite{jimenez-angeles:2004} Moreover, we have recently discussed a related situation, when two ionic electrolyte solutions consisting of large and small ions are in equilibrium with a thin film bounded by the flat semi-permeable membranes.\cite{vinogradova.oi:2012} As counterions could escape from electrolyte solutions, at some separation their clouds overlap and, as we have shown by the \emph{non-linear PB} theory and molecular dynamic simulations, give rise to a repulsive force between the membranes. However, many aspects of interactions of membranes have been given insufficient attention.

In the present paper, we are considering a particular case of spherical semi-permeable shells in osmotic equilibrium with \emph{outer} electrolyte solutions,
including solutions of multivalent ions.  We will use the
non-linear PB equation and integral equation theory based on the Ornstein-Zernike (OZ) equation with the hypernetted chain (HNC) closure to evaluate the ionic distributions, accumulated
charge inside the shell, and interaction between the shells.

\section{Model and Theory}

\begin{figure}
\centering \includegraphics[height= 4 cm]{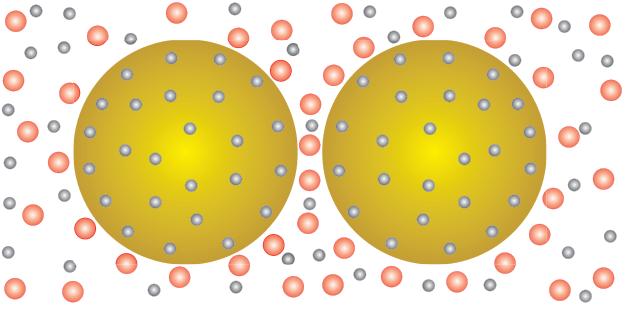}
\caption{Schematic of a system consisting of two shells immersed in a polyelectrolyte
solution. The shells (large spheres) are permeable for counterions (small spheres) only and impermeable for
polyions (medium-sized spheres). } \label{fig:capsule}
\end{figure}

We consider semi-permeable spherical shells of radius $R$ immersed in an electrolyte solution, as shown in Fig.~\ref{fig:capsule}. The electrolyte cations are characterized by charge $Z_+ e $, where $e$ is the electron charge, and bulk concentration $C_\infty ^+$, while the anions by $-Z_- e $ and $C_\infty ^-$, respectively. The shell is uncharged and infinitesimally thin. For the sake of definition we assume that the shell is permeable for cations and impermeable for anions. Furthermore, we impose electroneutrality of the bulk electrolyte, so that $Z_+ C_\infty^+ = Z_- C_\infty^-$.

\subsection{Nonlinear PB theory for single capsule in an electrolyte}

We first describe an isolated neutral shell immersed in an electrolyte using the Poisson-Boltzmann equation. We introduce the dimensionless potential $\varphi = \psi e / k_B T$, where $\psi$ is the electrostatic potential, $k_B$ is the Boltzmann constant, and $T$ the absolute temperature. The equations and the boundary conditions for the dimensionless potential are then given by
\begin{equation} \label{eq:8}
\begin{aligned}
& \varphi _i '' + \frac{2}{x} \varphi _i ' = -Z_+\left( \kappa R \right)^2 e^{-Z_+\varphi_i} \\
& \varphi _o '' + \frac{2}{x} \varphi _o ' = Z_+ \left( \kappa R \right)^2  \left( e^{Z_-\varphi_o} - e^{-Z_+\varphi_o} \right) \\
&  \varphi _i '\rvert _{x=0}=0 \ \  \ \ \ \ \ \varphi _i \rvert _{x=1} = \varphi _o \rvert _{x=1} \\
&  \varphi _o '\rvert _{x=\infty} =0 \ \ \ \ \ \ \varphi _i '\rvert _{x=1} = \varphi _o '\rvert _{x=1}
\end{aligned}
\end{equation}
where $x = \frac{r}{R}$, index $i$ refers to the capsule interior, while $o$ to the outer solution,
$\kappa^2 = 4\pi l_B C_{\infty}^+ $, $\kappa_o = \kappa \sqrt{Z_+\left( Z_- +Z_+\right) }$, and the Bjerrum length is $l_B = \frac{e^2}{4\pi\varepsilon_0\varepsilon k_B T}$.

To solve the equation we map the far field point $x = \infty$ to a finite $x^*$ and then, with the substitution
\begin{equation} \label{eq:9}
\xi_i = x_i \sqrt{x^* -1}, \ \ \ \ \xi_o = \sqrt{x^* -1} - \frac{x_o - 1}{\sqrt{x^* - 1}}
\end{equation}
we derive
\begin{equation} \label{eq:10}
\begin{aligned}
& \varphi_i '' +\frac{2}{\xi} \varphi ' _ i = - \frac{Z_+ \left( \kappa R\right)^2 }{x^* - 1} e^{-Z_+ \varphi _i} \\
& \varphi_o '' - \frac{2\sqrt{x^* - 1}}{ \left( \sqrt{x^* - 1} - \xi \right) \sqrt{x^* - 1} + 1} \varphi_o ' = \\
& \ \ \ \ \ \ \ \ \ \  =\left( x^* -1 \right) \left(\kappa R \right)^2 Z_+ \left(e^{Z_- \varphi _o} - e^{-Z_+ \varphi_o} \right)
\end{aligned}
\end{equation}
with the boundary conditions set by
\begin{equation} \label{eq:11}
\begin{aligned}
& \varphi_i ' \rvert _{\xi = 0} = 0, \ \varphi_i  \rvert _{\xi = \sqrt{x^* - 1}} = \ \varphi_o  \rvert _{\xi = \sqrt{x^* - 1}} \\
& \varphi_o ' \rvert _{\xi = 0} = 0,\ \left( x^* - 1\right)\cdot \varphi_i ' \rvert _{\xi = \sqrt{x^* - 1}} =- \ \varphi_o ' \rvert _{\xi = \sqrt{x^* - 1}}
\end{aligned}
\end{equation}
in the interval $\xi\in \left[ 0;\sqrt{x^* -1 }\right]$.

At $\xi = 0$ the first of the equations in (\ref{eq:10}) is singular. To remove the singularity we can use $ \varphi _i '' \rvert _ {\xi = 0} = - \frac{Z_+ \left( \kappa R \right) ^2 }{3 \left( x^* - 1\right) }e^{-Z_+\varphi_i\left( 0\right) }$. Numerical solution of simultaneous equations (\ref{eq:9}) can be then found using the following procedure. First, as we map an infinitely remote point onto a finite point $x^*$, we find $x_{\texttt{inf}}^*$  such that the potentials $\varphi _ i$ and  $ \varphi _o$ as well as their derivatives remain constant on the interval $\xi\in \left[ 0;\sqrt{x^* -1 }\right]$. The corresponding equations (\ref{eq:10}) with boundary conditions Eq. (\ref{eq:11}) can then be solved using the three-stage Lobatto IIIa formula (bvp4c function in Matlab).

In the limit of large radius or high electrolyte concentrations, $\kappa R \gg 1$, the shell is equivalent to a planar isolated semi-permeable membrane. The shell potential $\varphi_s$ is then defined by the boundary conditions $\varphi' |_{x=\pm {\infty}}=0$ and takes the form\cite{stukan.mr:2008,vinogradova.oi:2012}
\begin{equation} \label{eq:12}
\varphi_s^{*}=\frac{Z_+}{Z_-}\ln{\left( 1+\frac{Z_-}{Z_+} \right) }
\end{equation}
Thus, in monovalent electrolytes, the induced Donnan potential on the shell does not exceed $\varphi_s^{*}=\ln 2 \approx 0.693$. In a divalent electrolyte, its limiting value is $\varphi_s^{*} = \frac{1}{2}\ln 3 \approx 0.549$, and in trivalent electrolyte $\varphi_s^{*}=\frac{1}{3}\ln 4 \approx 0.462$. At room temperature these correspond to 18 mV, 14 mV, and 12 mV, respectively. Note that while the PB theory predicts the vanishing potential for larger polyion valencies, the mean-field model itself becomes increasingly inaccurate for multivalent ions. As the correlations come into play, the amount of charge the shell can accommodate will be even higher. In charged colloidal systems this leads to charge inversion, which, as we will see, appears also in our system. The charge inversion phenomenon is related to the charge discreteness and correlations, so that it lies beyond the mean-field description.\cite{grosberg.a:2002,boroudjerdi.h:2005}

We can calculate the accumulated charge inside the shell as $Q_n = Q/e=\left( \frac{R}{l_B} \right) \varphi_o'|_{x=1}$. A (neutral) shell with the accumulated inner charge will be referred below to as a (charged) capsule. We can now analyze an asymptotic behavior of the  external potential $\varphi_o'|_{x=1}$ at $\kappa R \rightarrow \infty$. It is easy to see that
\begin{equation} \label{eq:13}
\left. \left( \frac{d\varphi_o}{dx}\right) ^2\right|_{x=1} \approx \left( \kappa R \right)^2F\left[\varphi_s \right] + 4\left( \kappa R \right)\int\limits_0^{\varphi_s}\sqrt{F\left[ \varphi \right] } d\varphi
\end{equation}
where we introduced a function $F\left[ \varphi \right] = 2\left( \frac{Z_+}{Z_-}\left[ \exp{\left( Z_-\varphi\right) }-1 \right] + \left[  \exp{\left( -Z_+ \varphi\right) } -1  \right] \right) $. At $\kappa R\gg 1$ we have $\varphi_s \approx \varphi_s^{*}$, then in $\varphi_s^{*}$  we can omit terms of the order of $\left( \kappa R \right)^{-1} $. Similarly, we can omit the second term in Eq. (\ref{eq:13}) to find
\begin{equation} \label{eq:14}
Q_n^{*}=\kappa R \cdot \left( \frac{R}{l_B} \right) \sqrt{F\left[ \varphi_s^{*} \right] }
\end{equation}
In a symmetric electrolyte, $Z = Z_+=Z_-$, Eq. (\ref{eq:14}) takes a simpler form
\begin{equation} \label{eq:15}
Q_n^{*}=\kappa R \cdot \sqrt{8} \left( \frac{R}{l_B} \right)  \sinh{\left( \frac {Z \varphi_s^{*}}{2}\right) },
\end{equation}
which for a monovalent electrolyte reduces to $\frac{l_B Q_n^{*}}{R} = \kappa R = \kappa_oR/\sqrt{2}$. Thus, for large capsules or at high electrolyte concentrations the ion charge accumulated inside the shell is growing proportionally to the square root of the salt concentration ($\kappa$) and/or to the square of the radius.

\subsection{Interaction of capsules in the superposition approximation}

\begin{figure}
\centering \includegraphics[height = 6 cm]{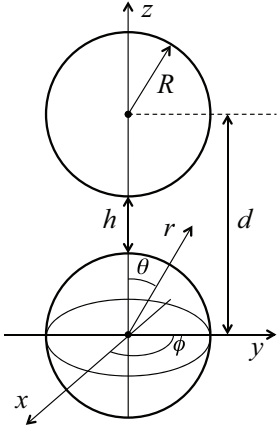}
\caption{Geometry of a system consisting of two interacting capsules. The polar angle $\theta$ and azimuthal angle $\phi$ in the spherical polar coordinates, the center-to-center distance $d$ and surface-to-surface separation $h$ are shown.} \label{fig:2capsules}
\end{figure}
In this section, we evaluate the interaction of two semi-permeable shells (capsules) in the superposition approximation where an analytical solution can be found.

We consider two spherical capsules of radius $R$ with their centers located at $\left(0,0,0 \right) $, $\left(0,0,d \right)$, as shown in Fig. \ref{fig:2capsules}, where $d$ is then the distance between the centers such that $d \geqslant 2R$. In the first-order approximation, the resulting potential will be a sum of the individual potentials of the shells\cite{stukan.mr:2008}
\begin{equation} \label{eq:2}
\varphi_o(x,y,z) = \varphi_o^{(1)}(x,y,z)+\varphi_o^{(1)}(x,y,z-d).
\end{equation}
This is justified provided the Donnan potential on the shell is small $\left( \varphi = \frac{e \psi}{k_BT}\ll 1 \right)$. This assumption is reasonable for a long-range part of the solution. In other words, our arguments are valid for a relatively large distance between the shells (see Appendix~\ref{A1} for more details).

For the shell centered at the origin, its surface potential can be rewritten in spherical polar coordinates as
\begin{equation} \label{eq:3}
\varphi_o^s(\theta,d) = \varphi_s + \frac{\varphi_s}{\vartheta(\theta,d) }\exp{\left( \kappa_o R\left( 1- \vartheta(\theta,d) \right) \right)}
\end{equation}
where $ \vartheta(\theta,d) = \sqrt{1+\left( \frac{d}{R}\right) ^2 - 2\left( \frac{d}{R}\right)\cos{\theta}}$ and the surface potential $\varphi_s$ is calculated numerically as described in Section 2.1. Then, the pressure exerted by the non-permeating ions on the shell will take the form
\begin{equation} \label{eq:4}
p(\theta,d) = \frac{Z_+}{Z_-}k_BT C_{\infty}^+\exp{\left( Z_- \varphi_o^s(\theta,d)\right) }
\end{equation}
The $z$-component of the net force acting on the shell at the origin is given by
\begin{equation} \label{eq:5}
(F_{\texttt{eff}})_z=\iint\limits_S p  \left( \textbf{n}_1 \cdot \textbf{n}_2 \right) d\sigma = - \iint\limits_S p \cdot \frac{z}{R} d\sigma
\end{equation}
where $\textbf{n}_1=-\left( x,y,z \right)/R $  is the unit normal vector to the capsule shell directed toward its center at $(0,0,0)$, $\textbf{n}_2=(0,0,1)$ the unit vector pointing from the center of capsule at $(0,0,0)$ toward the center of the second capsule at $(0,0,d)$. From Eqs. (\ref{eq:3}) and (\ref{eq:4}) we finally get
\begin{eqnarray}\label{eq:6}
 \nonumber
  F_{\texttt{eff}}(d) &=& R^2\int\limits_0^{\pi}\int\limits_0^{2\pi}p(\theta,d)\sin{\theta} \cos{\theta}  d\phi d\theta \\
   &=& 2 \pi R^2\int\limits_0^{\pi} p(\theta,d)\sin{\theta} \cos{\theta} d\theta
\end{eqnarray}

It is now instructive to evaluate the asymptotic behavior of the repulsive force $F_{\texttt{eff}}(r)$ for monovalent electrolyte in the long-distance limit, $r \gg R$. This can be done by expanding the exponential function in the integrand
\begin{equation} \label{eq:7}
e^{Z_- \varphi_o^s} = 1 + \frac{\varphi_s e^{\varphi_s} e^{\kappa_o R(1-r/R+ \cos{\theta} )}} {r/R} + O\left(\frac{r}{R}\right)^2 ,
\end{equation}
where we replaced $d$ with $r$ for consistency with the rest of the paper. This yields for monovalent electrolyte at $\kappa_oR \gg  1$
\begin{equation} \label{eq:7a}
F_{\texttt{eff}}(r)= \frac{2 \pi C_{\infty}^+ \varphi_s e^{\varphi_s} R^2 e^{-\kappa_o ( r - 2 R)}}{\kappa_o r} \approx \frac{\ln{(2)} \kappa_o R^2}{2 l_B} \frac{e^ { -\kappa_o ( r - 2 R) }}{r},
\end{equation}
An integration of the force Eq.(\ref{eq:7}) gives
\begin{eqnarray} \label{eq:7b}
\nonumber
\frac{U_{\texttt{eff}}(r)}{k_BT} & =&   \frac{2 \pi \varphi_s e^{\varphi_s} C_{\infty}^+ R^2 e^{2 \kappa_o R}}{\kappa_o} \Gamma(0,\kappa_o r) \\
 & \approx &  \frac{\ln(2)\kappa_o R^2 e^{2 \kappa_o R} }{2 l_B} \Gamma(0,\kappa_o r),
\end{eqnarray}
where $\Gamma(0,\kappa_o r)$ is the incomplete gamma function. In the following, we calculated $U_{\texttt{eff}}(r)$ numerically using this equation and integration of the mean-field force [Eq. (\ref{eq:6})].

We remark that our asymptotic results differ from what is expected for (impermeable) charged colloids. Indeed, we have obtained a Yukawa-like long-range behavior for the force (not for the potential), which indicates the regime of an ``ideal gas'' pressure of overlapping ionic atmospheres.\cite{Hoskin.mr:1955} Such a decay of the interaction force [Eq. (\ref{eq:7a}))] is likely a consequence of the two simplification used (a neglected effect of the second capsule on the charging and an assumption that $\kappa R \gg 1$). Nevertheless, as we will see below, this approach leads to quite accurate results at large separations.

\subsection{Integral equation approach}

In many experimental systems of interest multivalent ions are present. In this situation, effects of charge correlations on interactions between shells might be important. To address this issue, we need a more accurate treatment of the ionic distributions. Here we use the integral equation theory based on the OZ equation. To describe a capsule in an electrolyte, we will treat it as a separate type of particle, whose concentration is infinitesimally small. More generally, an inhomogeneous external field can be introduced as a new particle type at an infinite dilution.\cite{lozada-cassou:1981,lozada-cassou:1993,jimenez-angeles:2004}

A multicomponent OZ equation for $n+1$ particle types has the following form
\begin{equation} \label{eq:18}
h_{ij}(\mathbf r_{21})=c_{ij} (\mathbf r_{21}) + \sum\limits_{m=1}^{n+1} \rho_m \int h_{im}(\mathbf r_{23})c_{mj}(\mathbf r_{13})d \mathbf r_{3}
\end{equation}
where $\rho_m$ is the number density of particles of type $m$, $h_{ij}(\textbf{r}_{21})= g_{ij}(\textbf{r}_{21})-1$ and $c_{ij}( \textbf{r}_{21})$ the full and direct correlation functions for particle types $i$ and $j$ taken at $\mathbf{r}_2$ and $\mathbf{r}_1$,  $g_{ij}(\mathbf{r}_{21})$ is their pair distribution function, and  $\mathbf{r}_{21}=\mathbf{r}_2-\mathbf{r}_1$. The problem of calculation of the distribution functions from the known pair potential requires an additional relation for each pair $ij$, relating $h_{ij}(\mathbf{r}_{21})$ and $c_{ij}(\mathbf{r}_{21})$. Here we will use the hypernetted-chain closure, which is commonly used for electrolyte systems.\cite{jimenez-angeles:2004}
\begin{equation} \label{eq:19}
c_{ij} (\mathbf r_{21}) = -\beta u_{ij}(\mathbf r_{21}) + h_{ij}(\mathbf r_{21}) - \ln {g_{ij}(\mathbf r_{21})}
\end{equation}
where $u_{ij}(\mathbf{r}_{21})$ is the interaction potential for particles of types $i$ and $j$ and $\beta = 1/k_BT $ .

For an $n$-component fluid in an inhomogeneous field, the field can be considered as the $(n+1)$-th particle type (we denote it by index $\gamma$), whose density is infinitesimally small $\rho_\gamma \rightarrow 0$. The full correlation function for the field, type $\gamma$, and particle of type $j$ is given by
\begin{equation} \label{eq:20}
h_{\gamma j }(\mathbf r_{21})=c_{\gamma j} (\mathbf r_{21}) + \sum\limits_{m=1}^{n} \rho_m \int h_{\gamma m}(\mathbf r_{23})c_{mj}(\mathbf r_{13})d  \mathbf r_{3}
\end{equation}
The full correlation functions of the remaining particle types satisfy $n$-component OZ equations analogous to Eq. (\ref{eq:18}) but without $\gamma$ and alow us to calculate the direct correlation functions $c_{mj} (\mathbf r _{13})$. In this scheme, the pair correlation function $g_{\gamma j} (\mathbf r _{21})$ is equivalent to the inhomogeneous one-particle distribution function $g_j(\mathbf r_1)$ of particle type $j$ in an external field imposed by the capsule. Then, functions $h_{\gamma j}(\mathbf r _{21})$ and $c_{\gamma j}(\mathbf r _{21})$  can be replaced by $h_j (\mathbf r_1) = g_j (\mathbf r_1) - 1$ and $c_j(\mathbf r_1)$, respectively. Therefore, the local concentration of particles of type $j$ in the external field becomes
\begin{equation} \label{eq:21}
\rho_j (\mathbf r_1) = \rho _j g _j (\mathbf r_1)
\end{equation}
Now we can use the HNC closure (\ref{eq:19}) to substitute for $c_{\gamma j} (\mathbf r_{21})$ in Eq. (\ref{eq:20})
\begin{equation}\label{eq:22}
g_j(\mathbf r_1) =\exp \left\lbrace  -\beta u_j(\mathbf r_1)  + \sum \limits_{m=1}^{n} \rho_m  \int h_m(\mathbf r_3) c_{mj} (\mathbf r_{13} ) d\mathbf r_{3} \right\rbrace
\end{equation}
where index $\gamma$ is omitted for consistency with Eq. (\ref{eq:21}). Here, the function $c_{ij}(\mathbf r_{13})$ in the integral, Eq.(\ref{eq:22}) is replaced by the direct correlation function for $n$-component homogenous fluid, which is given by Eq. (\ref{eq:18}) without particles of type $\gamma$ via the HNC closure (\ref{eq:19}).

By solving Eq. (\ref{eq:22}) together with the closure relation (\ref{eq:19}), we can find $g_j(\mathbf r_1)$ and $c_j(\mathbf r_1)$. Hence, we can also calculate the distribution of particles $j$ near the shell, $\rho _j(r) = \rho_j g_j(r)$. Bulk concentration of particles of type $j$ is set by $C_{\infty}^{(j)} = \lim\limits_{r\rightarrow {\infty}} \rho_j g_j(r)$. Since $g_j(r) \rightarrow 1$ at $r \rightarrow \infty$, we have $C_{\infty}^{(j)}=\rho_j$. Similarly to Eq.(\ref{eq:22}) we can evaluate the pair distribution function for capsules $g_{\gamma\gamma} (\mathbf r_{21})$ in the limit $\rho_{\gamma}\rightarrow 0$ from
\begin{equation} \label{eq:23}
\ln{g_{\gamma\gamma}}\left( \mathbf r_{12} \right)  = -\beta u_{\gamma\gamma}(\mathbf r_{12}) + \sum\limits_{m=1}^{n} \rho_m \int h_{m}(\mathbf r_{23})c_{m}(\mathbf r_{13})d \mathbf r_{3}
\end{equation}
where $u_{\gamma\gamma}$ is the direct interaction potential for two capsules with their centers at $\mathbf r_1$ and $\mathbf r_2$. To calculate the potential of mean force for the capsules, we can use a one-component OZ equation (OCM) for a system containing capsules only such that it produces the same full pair correlation function as the full system, $h^{OCM}\left( \mathbf r_{12} \right)= h_{\gamma\gamma} \left( \mathbf r_{12} \right)$. In the limit $\rho _\gamma \rightarrow 0$, we find that $c^{OCM}\left( \mathbf r_{12} \right) = h^{OCM}\left( \mathbf r_{12} \right)$. The HNC closure (\ref{eq:19}) then gives $\beta U_{\texttt{eff}}(r)=-\ln{g_{\gamma\gamma}(r)}$, which is the required potential of mean force.

In the full system, we consider a two-component primitive electrolyte: positively charged spheres, which penetrate freely through the shell and mono or trivalent anions, which cannot penetrate into the capsule interior. The ion charge $Z_\pm e$  is put at the ion center. The interaction between cations as well as between cations and anions is set via a hard sphere potential. The solvent is considered as a continuous uniform dielectric medium characterized by the Bjerrum length $l_B=2$. Ions of types $i$ and $j$ separated by a distance $r$  between their centers interact via
\begin{equation} \label{eq:24}
u_{ij} (r)=
\begin{cases}
 \infty, \ \ \ \ \ \ \ \ \ \ \ \ \ \ \ \ \ \ \ \ \ \  \  r<a_{ij} \\
 l_B\left ( k_B T\right) \frac{Q_i Q_j}{r}, \ \ \ \ \  \ r \ge a_{ij}
\end{cases}
\end{equation}
where $i,j=+,-$, and $a_{ij}$ for monovalent ions are $a_{++}=a_{+-}=a_{--}= \sigma$, while for trivalent ions $a_{++}=a_{--}=\sigma$, $a_{+-}=2 \sigma$. Note that at low electrolyte concentrations the ionic distribution is sensitive only to $a_{+-}$, which determines the cation-anion closest approach distance and, hence, the binding energy. The two other diameters are unimportant since like-charged ions rarely collide at low concentrations due to Coulomb repulsion. As for the cation-anion distance, we have increased the contact distance to $a_{+-} = l_B$ to reduce the ion association, and make the correlation effects more pronounced. Global electroneutrality was provided by setting $Z_+ C_+ = Z_- C_- $. The interaction of anions with the capsule shell was described by the hard sphere potential
\begin{equation}\label{eq:25}
u_{\gamma -} (r)=
\begin{cases}
 \infty, \ \ \ \ \ \ \ \ \ \ \  r< R \\
 0,      \ \ \ \ \ \ \ \ \ \ \  r \ge R
\end{cases}
\end{equation}
We will further express all the length variables in $\sigma$.

We solved the multicomponent OZ equation with the HNC closure using the Ng's method with the modified version of program PLOZ.\cite{PLOZ}

\section{Results and Discussion}
\subsection{Distribution of charges and potentials near a single shell}

Now we focus on a distribution of ions and of electrostatic potentials around a single shell. First we analyze the predictions of the non-linear PB theory and its linearized version. Then we compare the results of the PB theory with calculations done using the integral equation approach.

\begin{figure}
\begin{center}
\includegraphics[width=0.9\linewidth]{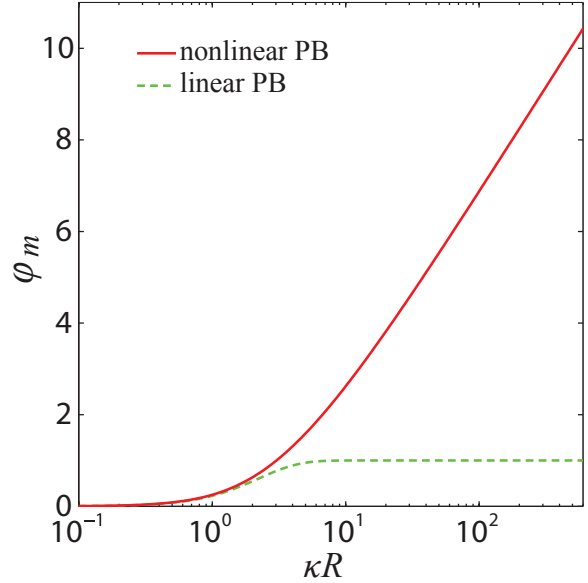}
\end{center}
\caption{Dimensionless electrostatic potential in the center of a capsule, $ \varphi_m$, for electrolytes with $Z_+=1$ and $Z_-=1$ obtained within the nonlinear PB equations (solid curve) and the linearized PB equations\cite{stukan.mr:2008} (dashed curve). We note that the curves for divalent and trivalent electrolyte, $Z_-=2,3$ would coincide with the result for monovalent salt on this scale and are omitted here. }\label{fig:fig3}
\end{figure}

The distribution of the electrostatic potentials at the center of the capsule, $ \varphi_m $,  versus $\kappa R$ is shown in Fig. \ref{fig:fig3}.  One can see that the potential at the center grows rapidly with $\kappa R$ and reaches very large values above $10 k_B T$ (or 250 mV) at $\kappa R \gg 1$ (no overlap of the inner ionic layers). The computed data show no dependence of $ \varphi_m $ on the valency of large ions, $Z_-$.  Note that the linear theory predicts a saturation of the potential at $\varphi_m = 1$ (see~\cite{stukan.mr:2008}) and, hence, significantly underestimates the potential at the center at large $\kappa R$. At $\kappa R \ll 1$ (strong overlap of the inner double layers), the potential at the center vanishes. Such a situation would be realistic for very dilute solutions and/or very small radius of the shell. These results are qualitatively similar to those obtained previously for a thin gap between two semipermeable membranes.\cite{vinogradova.oi:2012}

\begin{figure}
\begin{center}
\includegraphics[width=0.9\linewidth]{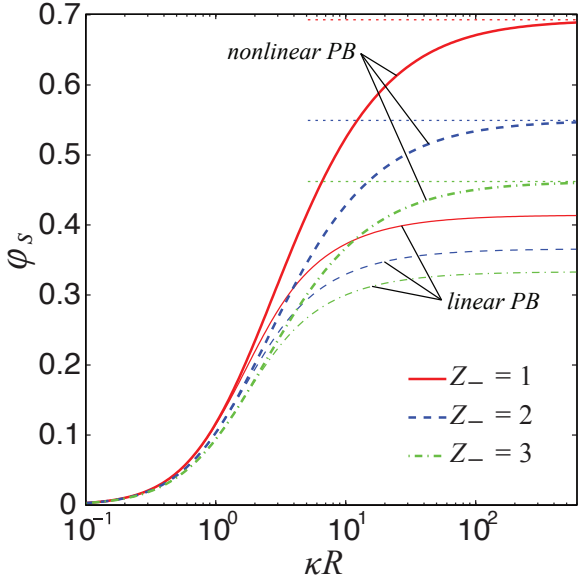}
\end{center}
\caption{Dimensionless shell potential, $\varphi_s(\kappa R)$, for electrolytes with $Z_+=1$ and $Z_-=1,2,3$ obtained from the nonlinear PB equations (thick curves) and the linearized PB equations (thin curves). The limiting Donnan potential of an isolated shell [Eq.(\ref{eq:12}] is shown by the dotted lines.} \label{fig:fig4}
\end{figure}

Fig.~\ref{fig:fig4} includes theoretical curves for the potential of the shell, $\varphi_s$, calculated for several $Z_-/Z_+$, by using the PB approach. At small $\kappa R$ all curves coincide, but at large $\kappa R$ the surface potential becomes sensitive to $Z_-/Z_+$ and decreases with this ratio, which is in agreement with the asymptotic result [Eq.(\ref{eq:12})]. Also included in Fig.~\ref{fig:fig4} are the theoretical curves predicted within the linearized PB theory. At large $\kappa R$ there is a discrepancy between predictions of nonlinear and linear theory, which becomes smaller at larger $Z_-/Z_+$. An explanation for for these deviations is the different (underestimated) asymptotic value of the bulk Donnan potential in the linear approach.\cite{vinogradova.oi:2012}

\begin{figure}
\begin{center}
\includegraphics[width=0.9\linewidth]{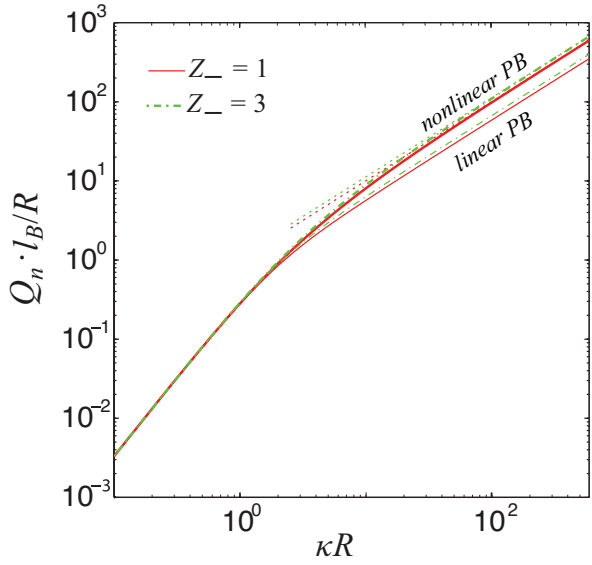}
\end{center}
\caption{Net charge inside the semi-permeable shell from the non-linear PB equation (thick curves), linearized PB equation (thin curves) and the asymptotic behavior (dotted curves) as predicted by Eq. (\ref{eq:14}) for an electrolyte with $Z_+=1$ and $Z_-=1,3$. }\label{fig:fig5}
\end{figure}

The build up of surface and center potentials is caused by a steric charge separation in our system. This is further illustrated by the data in Fig.~\ref{fig:fig5}, where the net accumulated charge inside the capsule is shown. The charge grows  linearly with $\kappa R$ when the latter is large, but is proportional to  $(\kappa R)^2$ when it is small. Note that the accumulated charge is not too sensitive to the valency of ions. The asymptotic law for the charge [Eq. (\ref{eq:14})] is also included (dotted-dashed curve), and becomes fairly accurate at large $\kappa R$.

\begin{figure}
\begin{center}
\includegraphics[width=0.9\linewidth]{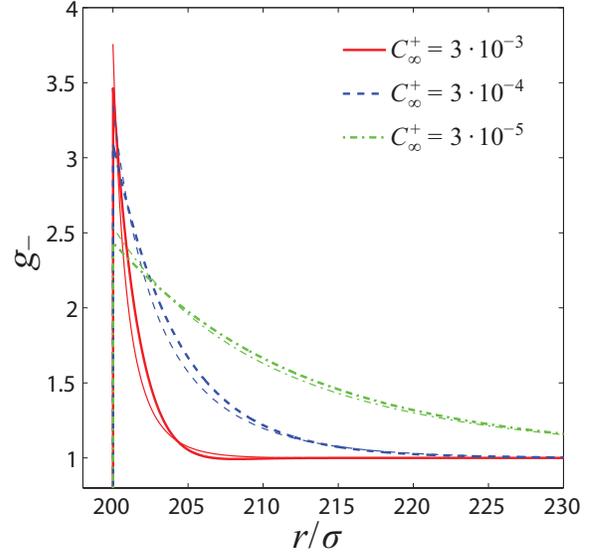}
\includegraphics[width=0.9\linewidth]{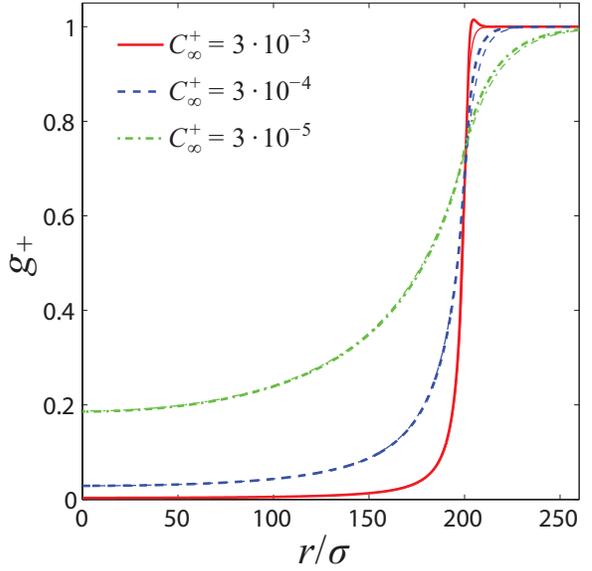}
\end{center}
\caption{Concentration profiles for (a) cations $(g_+(r))$, (b) anions $(g_-(r))$, obtained from the OZ-HNC calculations (thick curves) and PB equation (thin curves) for an electrolyte with $Z_+=1$ and $Z_-=3$.}\label{fig:fig6}
\end{figure}

Figure~\ref{fig:fig6} illustrates the ionic distributions, $g_-(r)$ and $g_+(r)$, in the vicinity of the shell, calculated  for several concentrations of trivalent electrolyte by using the integral equations technique. We see that at small concentrations the PB equation is very accurate, with only very small quantitative deviations from OZ-HNC. However, at large concentrations the agreement between the PB and OZ-HNZ is good only at very large $r$, but there are some qualitative discrepancies near the shell, both for anions and cations. The nonlinear PB approach leads to larger values of concentrations of
multivalent anions at the shell than those predicted by OZ-HNC. Integral equations theory also  shows oscillations of the local concentration of trivalent anions and
even counterions near the shell, which are an indication to ionic correlations in the system neglected in the mean-field  theory.  One can suggest that  the layer of polyanions at the shell formed due to their attraction to inner counterions is charged stronger than the interior, so that it condenses some outer cations. The PB theory fails to detect this phenomenon, which is similar to the effect of charge inversion for charged colloids.\cite{grosberg.a:2002,boroudjerdi.h:2005} The criterion for the charge inversion and like-charge attraction is normally given by the value of the coupling parameter $\Xi = 2 \pi \sigma_s l_b^2 Z^3 > 10$, where $\sigma_s$ is the surface charge density.\cite{boroudjerdi.h:2005} It is satisfied at high valencies and high surface charge densities. This approach can be used to quantify roughly the phenomenon we observe, but we suggest to use the inner charge, $\sigma_s = Q_n /(4 \pi R^2)$, to evaluate the onset of ion correlation effects. Then the $\Xi$ values for shells of radius $R=200$ and trivalent ions at the given range of concentrations varies from $\Xi \approx 1$ at the smallest to $\Xi \approx 23$ at the largest concentration, so that it is not surprising that we have observed the correlation effects in the latter case.

\subsection{Interaction between capsules}

\begin{figure}
\begin{center}
\includegraphics[width=\linewidth]{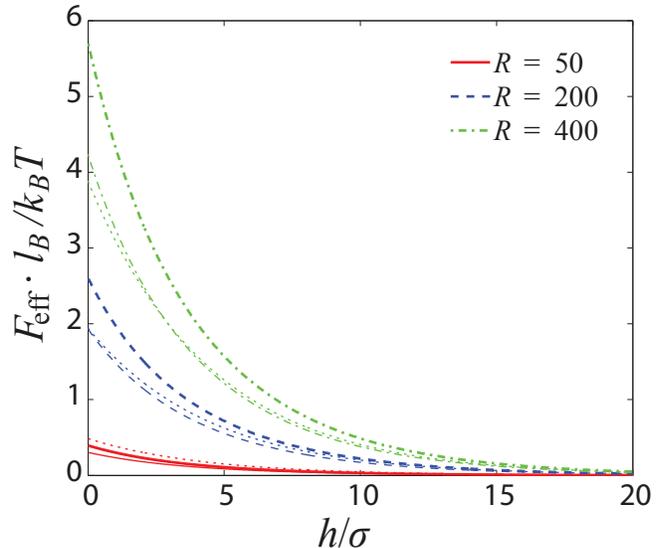}
\end{center}
\caption{Forces between capsules of radii $R=50$, 200, 400, immersed in a monovalent electrolyte of  concentration $C_{\infty}^+=10^{-3}$. The corresponding screening parameters are $\kappa_o R \approx 11, 45, 90$. Thick curves are OZ-HNC calculations, thin curves show the predictions of the mean-field theory [Eq. (\ref{eq:6})], and the asymptotic results [Eq. (\ref{eq:7a})] are shown by the dotted curves.}\label{fig:fig7}
\end{figure}

Fig.~\ref{fig:fig7} shows the force acting to the shell as a function of inter-shell separation, $h=r-2R$, calculated with OZ-HNC theory. The calculations are made for a monovalent electrolyte with fixed concentration by using several values of $R$. Also included are the predictions the mean-field [Eq. (\ref{eq:6})] and asymptotic [Eq. (\ref{eq:7a})] theory.  The capsule interaction is repulsive and its amplitude grows with $\kappa_o R$, in accordance with predictions of Eq.~(\ref{eq:7b}). At large $\kappa_o R$ the mean field theory visibly underestimates the value of a repulsive force. The striking result of our calculations is that approximate asymptotic solution is surprisingly accurate, even at very short distances.

\begin{figure}
\begin{center}
\includegraphics[width=\linewidth,clip]{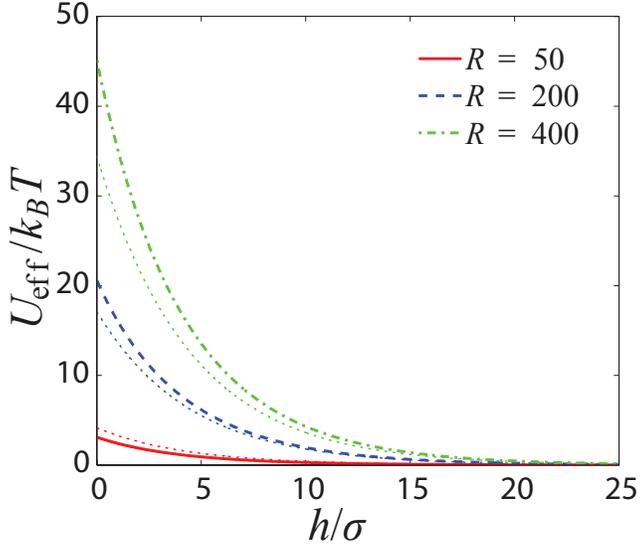}
\end{center}
\caption{Pair interaction potential for capsules of radii $R = 50$, 200, 400 in a monovalent electrolyte of concentration $C_{\infty}^+=10^{-3}$ at $l_B=2$ as obtained from the integral equation theory (thick curves) and the asymptotic formula (\ref{eq:7b}) (dotted curves). The corresponding screening parameters are $\kappa_o R \approx 11, 45, 90$.}\label{fig:fig9}
\end{figure}

The corresponding potential of mean force for capsules, $U_{\texttt{eff}}(r)$, is presented in Fig.~\ref{fig:fig9}.  Note that it reaches high values at a contact, especially for larger shell radii. For typical colloid systems these values would be sufficient to safely stabilize the dispersion of solid particles. Since the radii of shells correspond roughly to 20-150 nm, we conclude that suspensions of the shells of radii larger than ca. 100 nm should be stable against aggregation due to an accumulated inner charge. Fig.~\ref{fig:fig9} also includes calculations made by using an asymptotic formula, Eq.(\ref{eq:7b}). It can be again seen that it is surprisingly accurate and can be used to evaluate the interaction potential in many important situations.

\begin{figure}
\begin{center}
\includegraphics[width=\linewidth,clip]{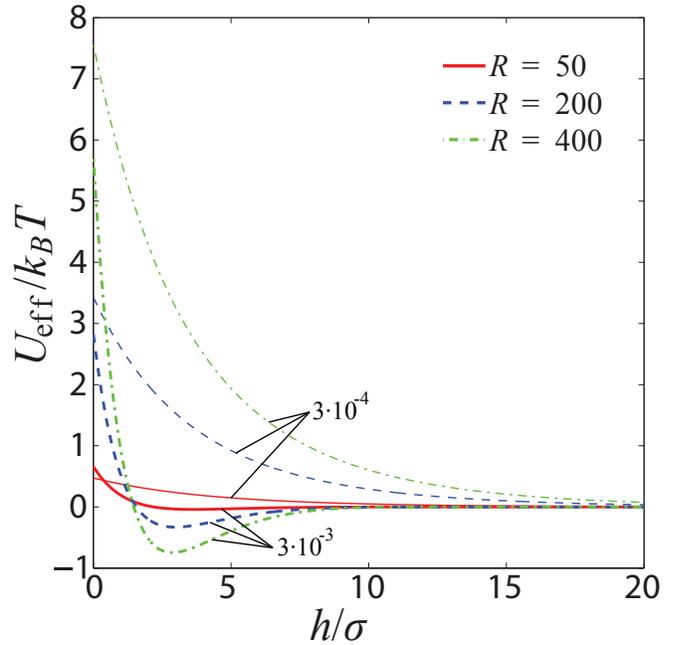}
\end{center}
\caption{Pair interaction potential for two capsules of radii $R=50, 200, 400$ in a trivalent electrolyte of concentration $C_{\infty}^+=3 \cdot 10^{-4} $ (thin curves) and  $3 \cdot 10^{-3} $  (thick curves) at $l_B=2$ as obtained from the integral equation theory. }\label{fig:fig10}
\end{figure}

The picture changes drastically in a solution of trivalent ions as it is seen in Fig.~\ref{fig:fig10}. At small electrolyte  concentration the interaction is similar to that in monovalent electrolyte. However, at large concentrations the interaction changes qualitatively: an attractive force appears at intermediate distances. The criterion for like-charge attraction is the same as for the apparent shell charge inversion. Note that the range of the attraction is the same for all capsule sizes and reflects the mean distance between the polyions within the ``condensed'' layer.\cite{boroudjerdi.h:2005} We also remark that the attraction becomes stronger for larger capsules.

The phenomenon of like charge attraction of semipermeable  shells is similar to that observed earlier for charged colloids. It had been reported originally\cite{guldbrand.l:1984,rouzina.i:1996} that the attraction arises only at certain concentration of the multivalent salt. Later it has been shown that this  effect is generic and observed in mixtures of charged colloids and polyelectrolytes.\cite{walter.hw:1996,bouyer.f:2001,grosberg.a:2002} As we discussed above, the potential of the shell decreases with the polyion charge (see Eq. (\ref{eq:12})). At the same time, the energy of electrostatic polyion binding at the surface, is growing as $Z_- \varphi_s \approx  \ln Z_-$. Thus, the polyions will be stronger attracted to the shell, so that the ionic correlation criteria are easier to satisfy at the higher polyion charge.  This was indeed found in the recent simulation study.~\cite{stukan.mr:2008} Hence, we expect a very pronounced attraction between the shells immersed into polyelectrolyte solutions and the aggregation of the shells as a result.

\section{Conclusions}

We have presented the nonlinear PB theory, including numerical calculations and asymptotic analysis, and OZ-HNC data on interaction of semipermeable shells
immersed in an electrolyte solution. We found that electrolyte-mediated interaction between the shells in low valency electrolytes is always repulsive and is sufficient to stabilize their dispersion. However, it can become attractive due to ion correlation effects already for a trivalent electrolyte provided its concentration is large enough.

\section*{Acknowledgement}

This research was partly supported by the Russian Foundation for Basic Research (grant 12-03-00916).

\appendix
\section{Appendix: Asymptotic form of the electrostatic potential at large distances}\label{A1}

Here we show that an accurate solution can be found if we solve the nonlinear equation for a shell potential, but use its asymptotic form, which is similar to the solution of the linearized PB equation~\cite{stukan.mr:2008}
\begin{equation} \label{eq:1}
\varphi^{(1)}_o = \varphi_s \frac{R}{r}\exp{\left( - \kappa_o\left(r - R \right)  \right)}.
\end{equation}
It is known that the far field part of the solution of the nonlinear PB equation for spherical geometry can be presented in the same functional form. In Refs. \cite{dukhin:1970,dukhin:1970a}, it has been shown that the potential for charged colloidal particles calculated from the nonlinear PB equation can be presented at large distances as
\begin{equation} \label{eq:16}
\varphi\left( r \right) = \frac{B}{r} \exp{\left[ -\kappa_o\left(r - R \right) \right] }
\end{equation}
where the coefficient $B$ is given by
\begin{equation} \label{eq:16a}
\begin{aligned}
& B = 4 \tanh{ \left( \frac{\varphi_s}{4}\right) } \left[ 1 + \frac{1}{2\kappa_oR} \tanh{\left( \frac{\varphi_s}{4}\right)}  \right .  \\
& \left . + \frac{1}{16 \kappa_o R} \int_0^{\varphi_s} F(f) \frac{f_1^2(f)\sinh{(f)}}{\kappa_oR - \ln { \frac{\tanh{(f/2)}}{\tanh{(\varphi_s/4)} } }} df  \right]
\end{aligned}
\end{equation}
\begin{equation} \label{eq:16b}
f_1 = \sinh{ \left( \frac{\varphi_s}{2}\right) } \left[ \tanh^2{ \left( \frac{\varphi_s}{2}\right)} - \tanh^2{ \left( \frac{f}{2}\right)} + 2 \ln { \frac{\tanh{(f/2)}}{\tanh{(\varphi_s/4)} } } \right]
\end{equation}
\begin{equation} \label{eq:16c}
F(f) =   \frac{\cosh{(\varphi_s/2)}}{\sinh^2{ (\varphi_s/2) }}  -  \frac{\cosh{(f/2)}}{\sinh^2{ (f/2) }} + \ln { \frac{\tanh{(f/4)}}{\tanh{(\varphi_s/4)} } }
\end{equation}
A similar approach for calculation of effective interactions between charged colloids has been also used in  the limit $\kappa R \to \infty$.\cite{aubouy.m:2003} This procedure is associated with colloidal charge renormalization, as it changes the prefactor of the screened Coulomb potential. In our system, however, the potential is small, so that the second and the third terms of Eq. (\ref{eq:16a}) can be safely neglected, so that $B$ is very close to $\varphi_s$. In other words, the use of the potential in the form Eq. (\ref{eq:1}) is well justified.

\footnotesize{
\bibliography{capsule}
\bibliographystyle{rsc}
}
\end{document}